# Cycles of Activity in the Jovian Atmosphere

L.N. Fletcher

Department of Physics & Astronomy, University of Leicester, University Road, Leicester, LE1 7RH, UK; leigh.fletcher@leicester.ac.uk

**Abstract (100 words)**

Jupiter's banded appearance may appear unchanging to the casual observer, but closer inspection reveals a dynamic, ever-changing system of belts and zones with distinct cycles of activity.  Identification of these long-term cycles requires access to datasets spanning multiple jovian years, but explaining them requires multi-spectral characterization of the thermal, chemical, and aerosol changes associated with visible color variations.   The Earth-based support campaign for Juno's exploration of Jupiter has already characterized two upheaval events in the equatorial and temperate belts that are part of long-term jovian cycles, whose underlying sources could be revealed by Juno's exploration of Jupiter's deep atmosphere.

**Main Text (2000 words)**

Four decades of robotic planetary exploration, combined with an ever-expanding record of Earth-based remote observation, have revealed astounding insights into the atmospheric phenomena shaping the giant planet Jupiter, but have also raised many unanswered questions [Ingersoll et al., 2004; Vasavada and Showman, 2005].  NASA's Juno mission, in orbit around Jupiter since July 2016, is set to transform our understanding of jovian processes that are largely inaccessible to Earth-based observers – in the polar regions, the magnetospheric environment, and the composition and structure of the planet hidden deep below the visible clouds.  But compared to Jupiter's 11.9-year orbital cycle, Juno's orbital exploration will be comparatively short-lived.  Long-term Earth-based records of atmospheric variability (Fig. 1) therefore provide essential temporal context to Juno's new discoveries.  In addition to monitoring Jupiter's ever-changing appearance, an international campaign of Earth-based professional and amateur support for Juno is providing both spatial context for Juno's close-in views, and spectral support in wavelengths inaccessible from Juno's remote sensing experiments.  This support campaign has both characterized Jupiter's atmospheric state (belt/zone structure, winds, and storms) at the start of Juno's mission, and yielded discoveries in its own right, as reported in this Juno special collection for GRL.

**Time variability of Jupiter's banded structure**

Jupiter's characteristic banded structure of bright white zones and darker belts can evolve over hours (convective storm eruptions), months and years in response to poorly-understood meteorological activity within (and potentially below) the main cloud decks. These 'upheavals' in the banded structure appear to be most severe within the cyclonic, darker belts (Fig. 1), with patterns of variability observed in both the Equatorial and Temperate Belts, albeit with ill-defined and irregular periodicities (Table 1). These profound changes encircle the whole planet and provide valuable insights into jovian atmospheric dynamics, cloud and chemical structure. Notable modes of jovian belt/zone variability in Table 1 can be subdivided into three broad categories: (a) gradual coloration episodes (reddening) that could be related to aerosol changes via photochemical processing of newly-advected material; (b) gradual whitening caused by condensation/thickening of aerosol layers (known as fades), or darkening caused by evaporation/thinning of condensate aerosols (revivals); and (c) sudden outbreaks of convective plumes [Stoker, 1986, Gierasch et al., 2000], both on jetstreams and within major belts, leading to some revival events.

This latter category can be extreme, to the delight of Earth-bound astronomers – the South Equatorial Belt (SEB), the brown band immediately south of Jupiter's equator, undergoes significant whitening events (fades) over many months, before vigorous convective activity 'revives' the typical brown appearance [Rogers, 1995; Sanchez-Lavega et al., 1996; Sánchez-Lavega & Gomez, 1996]. Such strong moist convective activity may be associated with jovian lightning [Little, 1999]. During these fade events, the SEB can appear hidden and almost indistinguishable from the Equatorial Zone (EZ) to the north and the South Tropical Zone (STrZ) to the south (Fig. 1). The last SEB fade and revival occurred between 2009 and 2011 [Fletcher et al., 2011, Perez-Hoyos et al., 2012, Fletcher et al., 2017], and assessment of the historical record shows that these dramatic cycles sometimes occur at intervals of about 3 years, but periods between fades can be as long as 36 years, with many intervening years of 'normal' SEB activity. If Juno had been observing Jupiter during one of these events, as Pioneer 10 and 11 did, we might expect aerosol distributions (measured by JunoCam and JIRAM) and gaseous distributions (e.g., ammonia measured by the microwave radiometer) to be anomalous compared to the long-term record.

**Jupiter's Belts in 2016**

Variability within the other Jovian belts (Table 1) can be similarly dramatic, and the 2016-17 campaign of Juno-supporting observations has witnessed events at two distinct locations that form part of this wider cycle of Jovian atmospheric variability. Jupiter's North Equatorial Belt (NEB), which typically resides between 7-17°N latitude, showed signs of expanding northwards into the neighboring North Tropical Zone (NTrZ) during 2015-16. Fletcher et al. [this issue] describe this as part of a 3-to-5-year cycle of NEB activity that has occurred since 1988, temporarily removing white aerosols from the NTrZ and giving the appearance of an expansion of the brown NEB. However, the NEB had receded to its normal latitudinal extent by the time of Juno's arrival, making the planet appear 'normal' at the start of Juno's orbital exploration. What came next was even more spectacular – Sánchez-Lavega et al. [this issue] report the eruption of four extraordinary plumes near Jupiter's fastest eastward jet at 23.9°N, characteristic of another cycle of atmospheric variability in the North Temperate Belt (NTB) [Sánchez-Lavega at el., 1991, 2008]. The October 2016 NTB eruption was remarkable for several reasons: (i) Juno had just arrived in the jovian system to observe the phenomenon; (ii) four plumes at widely separated longitudes were identified; and (iii) the Earth-based support campaign was able to provide remarkable temporal coverage even though Jupiter was still relatively close to solar conjunction.

Unlike the less organized NEB expansion events, these NTB disturbances are likely powered by moist convective plumes rising from the depths of Jupiter's water cloud [Sánchez-Lavega et al., 2008], bearing many similarities to the processes controlling SEB revivals.  The rising plumes interact with the background zonal flows to form characteristic shapes [Hueso et al., 2002] – whereas SEB revivals take on an 'S-shaped' form due to their location between an eastward jet to the north and a westward jet to the south [Sánchez-Lavega and Gómez, 1996], the NTB plumes moved eastward faster than the ambient zonal flows, generating wakes of bright and dark spots that went on to encircle the whole latitude band.  The evolution of these disturbances is well reproduced in numerical simulations by injecting heat pulses or mass sources to represent convective plumes [Sánchez-Lavega et al., this issue].  The 2016 plumes appeared to erupt from the anticyclonic region just to the south of the eastward zonal jet at 23.9°N.  This is technically the northern edge of the NTrZ, but numerical simulations by Sánchez-Lavega and colleagues suggest that the disturbance morphology can only be correctly reproduced if the plumes originate from the cyclonic NTB north of the zonal jet.  Intriguingly, this suggests that the plumes experience a significant tilt equatorward as they ascend, such that their cloud-top appearance near 22.2-23.0°N doesn't represent their deep sources near 24.5-25.0°N.

Furthermore, the 2016 NTB eruption stands out in that four individual plumes were involved, at latitudes that were geographically separated from one another.  What process could trigger plumes in these four different locations over a period of just 2-3 weeks?  This hints at some connection between these plume sites at deeper unobserved levels below the visible clouds.  Could some deep atmospheric wave pattern or instability be responsible?  Or could multiple hidden meteorological features, propagating at depth, be responsible for destabilizing the weather layer to convective outbursts?  SEB revival plumes appear to erupt within cyclonic circulation patterns known as barges [Fletcher et al., 2017], whose low-pressure centers may serve to raise isentropes in deeper layers and possibly cause the upwards displacement near the water clouds necessary to initiate convection [Dowling and Gierasch, 1989].  These SEB barges were practically invisible at the time of the SEB plume eruptions in 2010 but were likely still present at deeper levels.  No comparable cyclonic circulations were noted in the NTB jet prior to the eruption.  However, it has been noted that the zonal jet at 23.9°N gradually accelerates prior to these disturbances [Sánchez-Lavega et al., 2008; Rogers et al., 2006], and the same was true in 2016 [Hueso et al., this issue].

**Cyclic activity in the North Temperate Belt**

The deep environmental changes responsible for this jet acceleration remain unclear, but by identifying these precursors, Jupiter researchers are able to provide crude predictions of when an NTB outbreak might be imminent.  Rogers (1995) presents a historical overview of NTBs outbreaks, and we note that the NTB has been through this cycle at least six times since spacecraft observations began – in 1975 [Sánchez-Lavega and Quesada, 1988], 1980, possibly 1985 during solar conjunction, 1990 [Rogers et al., 1995; Sánchez-Lavega et al., 1991], 2007 [Sánchez-Lavega et al., 2008] and in 2012 when Jupiter was in solar conjunction [Rogers, 2012].  Thus, successive NTB outbreaks can occur at intervals of 5 years or longer.

If the individual plumes themselves are hard to study because of their short lifetimes, the aftermath of these events is sometimes easier to characterize.  The 2016 NTB disturbance produced a new, red band over the southern half of the NTB (22.8-26.7°N), whereas the northern half of the NTB became gray and turbulent up to the westward jet on its northern edge (near 32°N).  This reddening of the southern NTB has been referred to as an NTB revival [Rogers, 1995], as this belt is often 'faded' before the outbreak and dark gray at other times, as can be seen in Fig. 1 after the 2007, 2012 and 2016 outbreaks.  However, during both SEB revivals [Fletcher et al., 2017] and NEB expansion events [Fletcher et al., this issue], low-albedo red-brown regions are typically associated with high emission at 5-µm (and other thermal-infrared wavelengths), suggestive of a removal of aerosol opacity to allow radiance to escape from deeper, warmer levels of the atmosphere.  The reddened NTB appears to be different, both in the vivid nature of the red color, and the fact that it has no 5-µm bright counterpart.  This combination of red color and high 5-µm opacity is also observed in Jupiter's Great Red Spot, raising the possibility of a photochemical explanation for this new band.  For example, material dredged upwards by the erupting plumes is exposed to solar ultraviolet radiation, which may lead to a chain of photochemical pathways [e.g., Carlson et al., 2016] to produce a high-altitude haze containing a red coloring agent.  SEB revival events tend to take on an orange coloration during their latter stages [Fletcher et al., 2017], and the Equatorial Zone periodically takes on a orange/yellow appearance (Fig. 1, Table 1), and we speculate that similar photochemical modification of newly-emplaced material could be at work.  The existence of this reddened NTB band, which persists for some years before it narrows and fades away, could help to improve our understanding of the NTB life cycle.  For example, the historical record of Rogers (1995) shows a 5-year activity cycle for this belt, consistent with recent events in 2007, 2012 and 2016.

**Identifying the causes of jovian variability**

Several different processes appear to be at work during these cyclic events on Jupiter: eruption of moist convective plumes that interact with the ambient winds in the SEB and NTB; large-scale subsidence and cloud clearing in the SEB and NEB; and production of orange-red aerosols (potentially via photochemistry) following NTB outbreaks and SEB revivals.  However, events at particular latitudes are rarely identical, and with only ~3 jovian years of high quality digital observations (since Pioneer in 1973-74) we must be cautious about generalizations.

For each cycle discussed above, a combination of improved observations and numerical simulation is required to address some of the outstanding questions.  What controls the time intervals between successive disturbances or expansion episodes?  The small axial tilt of Jupiter (3°) means that seasonal oscillations are unlikely to play a role, but the 4.4-year cycle of Jupiter's quasi-quadrennial oscillation (QQO, [Leovy et al., 1991]) that shapes the equatorial stratosphere will have secondary circulation patterns in the extra-tropics.  Further work is needed to determine whether this could provide a cyclic influence on the upper tropospheric temperatures [e.g., Simon-Miller et al., 2007], and potentially a connection to the aerosol changes observed in visible light.  The next generation of numerical models, incorporating the active meteorology of jovian water [e.g., Sugiyama et al., 2014, Li and Ingersoll, 2015], might hold the key to the buildup of convective available potential energy preceding an outburst.  What determines the longitude for the first plumes of an SEB revival, or the first regions of the NEB to expand?  Why does an NTB disturbance erupt on the peak of the prograde jet, and how can it erupt almost simultaneously from multiple, well-separated longitudes?  Analysis of the first microwave observations from Juno, along with Earth-based observations in the millimeter and radio regimes [e.g., de Pater et al., 2016], could reveal interesting insights into the deep atmospheric phenomena underlying these belts.  What determines the lifetime of the convective plumes, and the post-plume evolution of the belts (e.g., the aerosol clearing of an SEB revival, the reddening of the NTB)?

Perhaps the most intriguing question of all: are any of these cyclic events in different jovian domains somehow connected as part of an atmospheric cycle that is global in scale [Rogers, 1995]?  And if so, what mechanism (waves, general circulation, etc?) permits the transmission of energy and information between adjacent latitude bands?   Or are the changes purely stochastic, governed by the unpredictable dynamical changes and instabilities at smaller scales?  After all, correlation does not imply causation.  Nevertheless, with continued improvements to our observing and image processing capabilities, as well as access to infrared observations to characterize the chemistry, temperatures and cloud properties coinciding with these visible-light changes, the time series for studying Jupiter's cyclic variations is becoming ever more comprehensive.

**Acknowledgements**

Fletcher is supported by a Royal Society Research Fellowship and a European Research Council Consolidator Grant at the University of Leicester.  I thank Dr. John Rogers and Dr. Santiago Pérez-Hoyos for their thorough reviews and discussion of the concepts in this commentary.  The data used in Figure 1 are available from the primary author and are listed in Fletcher et al. [this issue] and Rogers (2017).

| Band | Name | Latitude Range | Event(s) | Description |
|---|---|---|---|---|
| **NTB** | North Temperate Belt | 24-31°N | Jet outbreaks and revival | Outburst of convective plumes on southern jet; reddening of southern part of belt (last in 2016, Sánchez-Lavega et al., this issue). |
| **NEB** | North Equatorial Belt | 7-17°N | Expansion events | Narrowing and broadening into neighboring zone with 3-5 year cycle (last in 2015-16, Fletcher et al., this issue). |
| **EZ** | Equatorial Zone | 7°S-7°N | EZ coloration episodes | Orange/yellow tint in EZ (last in 2012-13) |
| **SEB** | South Equatorial Belt | 7-20°S | Fade and revival cycles | Whitening, then convective revival (last in 2010-11). Smaller SEB outbreaks occur during 'normal' state. |
| **STB** | South Temperate Belt | 27-32°S | Fade and revival in discrete sectors | Outburst of spots on northern jet, but none as large as NTB outbreaks (Rogers, 1995). |

**Table 1:** Characteristic variability in Jupiter's tropical and mid-latitude belts (the equatorial zone is also included). Latitude ranges are taken from Cassini-derived jet locations [e.g., Porco et al., 2003]. Events in the NTB and NEB are discussed in this GRL special collection. For historical overviews of other events, the reader is referred to Rogers (1995).

**Figure 1:** Jupiter's belt/zone variability observed between the 2000 Cassini flyby and the Juno observations in 2017. Representative longitude ranges were extracted from maps of amateur observations compiled by Rogers (2017) (see Fletcher et al., this issue, for a complete list of observers and dates). Several examples of NTB revivals, NEB expansions, EZ coloration episodes and SEB revivals can be seen during this 17-year (1.5-Jupiter-year) time span. Exploring this variability with (i) a broad wavelength range extending into the infrared; (ii) higher frequency of observations within a particular apparition; and (iii) extended time coverage in the 20[th] century is a goal of future Jovian research.

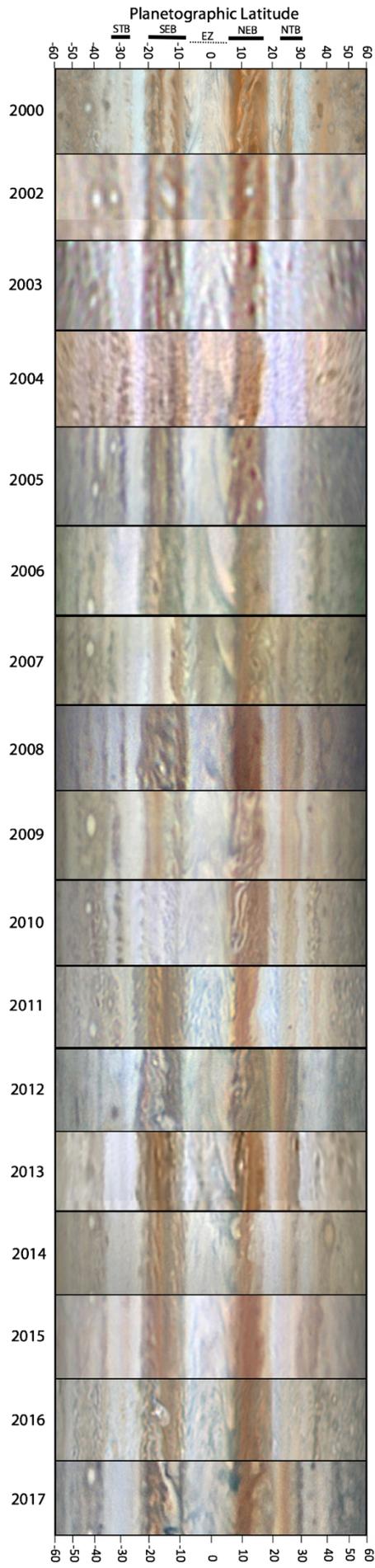